\documentclass{raa}

\usepackage{graphicx,times} %for PS/EPS graphics inclusion, new
\usepackage{graphicx}
\usepackage{amsmath}
\usepackage{lineno}
\usepackage{url}
\usepackage{array}
\usepackage{pifont}
\usepackage{caption}
\usepackage{natbib}
\usepackage{amssymb,amsmath}
\usepackage{tablefootnote}
\usepackage[colorlinks,unicode=true,linkcolor=green,anchorcolor=red,citecolor=blue,filecolor=red,pagecolor=red,urlcolor=red]{hyperref}

\bibpunct{(}{)}{;}{a}{}{,}

\begin{document}

\title{Observation data pre-processing and scientific data products generation of POLAR}
\volnopage{Vol.0 (20xx) No.0, 000--000}
\setcounter{page}{1}
\author{
Zheng-Heng~Li\inst{1,2}
\and Jian-Chao~Sun\inst{1}
\and Li-Ming~Song\inst{1}
\and Bo-Bing~Wu\inst{1}
\and Lu~Li\inst{1}
\and Xing~Wen\inst{1,2}
\and Hua-Lin~Xiao\inst{1,3}
\and Shao-Lin~Xiong\inst{1}
\and Lai-Yu~Zhang\inst{1}
\and Shuang-Nan~Zhang\inst{1}
\and Yong-Jie~Zhang\inst{1}
}

\institute{Key Laboratory of Particle Astrophysics, Institute of High Energy Physics (IHEP), Chinese Academy of Sciences, Beijing 100049, China; {\it Zhengheng.Li@ihep.ac.cn} \\
\and University of Chinese Academy of Sciences, Beijing 100049, China \\
\and Paul Scherrer Institut, 5232 Villigen PSI, Switzerland \\
\vs\no
{\small Received~~2018 August 14;\quad accepted~~2019 January 3}
}

\abstract{
POLAR is a compact space-borne detector initially designed to measure the polarization of hard X-rays emitted from Gamma-Ray Bursts in the energy range 50-500\,keV. This instrument was launched successfully onboard the Chinese space laboratory Tiangong-2 (TG-2) on 2016 September 15. After being switched on a few days later, tens of gigabytes of raw detection data were produced in-orbit by POLAR and transferred to the ground every day. Before the launch date, a full pipeline and related software were designed and developed for the purpose of quickly pre-processing all the raw data from POLAR, which include both science data and engineering data, then to generate the high level scientific data products that are suitable for later science analysis. This pipeline has been successfully applied for use by the POLAR Science Data Center in the Institute of High Energy Physics (IHEP) after POLAR was launched and switched on. A detailed introduction to the pipeline and some of the core relevant algorithms are presented in this paper.
}

\authorrunning{Z.H. Li et al.}
\titlerunning{Data Pre-Processing \& Data Products of POLAR}

\maketitle

\section{Introduction}\label{sec:introduction}

POLAR~(\citealt{PRODUIT2005616}) is a compact space-borne detector which is specially designed and dedicated to measuring the polarization of hard X-ray/Gamma-Ray emissions from Gamma-Ray Bursts (GRBs) in the energy range 50-500\,keV. POLAR uses plastic scintillators (PSs) as the main detection material along with the multi-anode photomultiplier tube (MAPMT) to measure the polarization by recording the distribution of azimuthal scattering angle of the incoming photons interacting with the detection material through the process of Compton scattering~(\citealt{Xiong2009}). The POLAR instrument consists of two major components, the Outer BOX mounted on the outer surface of Tiangong-2 (TG-2) (OBOX) and the Inner BOX mounted in the interior of TG-2 (IBOX), the details of which are described in \citealt{Produit2017}. POLAR was launched onboard the Chinese space laboratory TG-2 on 2016 September 15 and switched on successfully a few days later. Then, a series of in-orbit tests were performed and the normal science data acquisition was started subsequently, during which time tens of gigabytes of detection raw data were produced in-orbit by POLAR and transferred to the ground every day. The directly downloaded raw data from POLAR consist of three types, which are SCI\_0B, AUX\_0B and ENG\_0B. The raw data of type SCI\_0B contain the scientific detection data of POLAR and the engineering data are stored in the raw data of type AUX\_0B and ENG\_0B. All the raw data from POLAR are in the form of a data stream of binary packets generated by the Field-Programmable Gate Array (FPGA) which are not easy to analyze directly for science purposes. Therefore, pre-processing all the detection raw data and generating well-organized, high level scientific data products as quickly as possible are the first important steps and are required by the POLAR experiment for the following science analysis, such as data quick-look, in-orbit calibration~(\citealt{LI20188}), polarization analysis for the detected GRBs, etc. The requirements of data pre-processing and scientific data products generation for POLAR mainly include:
\begin{itemize}
\item iterating and decoding data packets as well as converting engineering data from binary values to physical values;
\item doing time alignment between packets from the Center Trigger computer (CT) and the Front-End Electronics (FEE) in modules for each event;
\item reconstructing the absolute time of each event;
\item doing the coordinate transformation to calculate the direction of POLAR in celestial coordinates (J2000);
\item doing proper data merging and splitting to organize all different kinds of data into a simple data structure which is friendly for science analysis.
\end{itemize}

A full pipeline and the related software to automatically do these works were designed, developed and tested before the launch date of POLAR and fully verified during the in-orbit test phase of POLAR after launch. After the raw data of POLAR arrive at the Institute of High Energy Physics (IHEP), the three types of raw data are firstly decoded separately, and for the scientific detection data the time alignment for physical events are also performed in the decoding phase. Then using the decoded engineering data the absolute time of each event can be reconstructed. Finally, for convenience of scientific analysis, all of the different kinds of data are merged together with a simple data structure, keeping only the necessary data fields that are directly needed by science analysis. The merged data are called SCI\_1Q and are used as the input and standard data format for the science data analysis pipeline. Based on the data product of SCI\_1Q, some higher scientific data products for publication can also be generated concerning the data archiving and scientific data products of some specific celestial events like GRB and solar flare events.

As a part of the basic software suite of the POLAR Science Data Center (PSDC) at IHEP, this pipeline has been successfully running and providing service for the science data analysis of researchers after POLAR was launched. This pipeline can work automatically without manual intervention and will send an email to notify the researchers that new scientific data products have been generated. Up to now, with scientific data products generated by the pipeline of the current version, the in-orbit instrument performance study and calibration have been finished and are presented in \citealt{LI20188}, and the first scientific results of POLAR have been produced and published in \citealt{Zhang.Kole.2019} recently. This paper will give a detailed introduction of this pipeline and the main algorithms that are important and specially needed by POLAR. The programming languages C++ and Python are widely used and the ROOT\footnote{\url{https://root.cern.ch}} Library developed by CERN is the main tool that was employed in the pipeline to organize the scientific data products. Besides the pipeline, a table describing the main data structure specification for the high level scientific data products of POLAR in ROOT files will also be presented in this paper.

\section{Raw data from POLAR}\label{sec:raw_data}

The scientific detection data and engineering data from POLAR are firstly collected online by and temporarily stored on the TG-2 platform, then transferred via telemetry to the ground station where all the data will be transferred to the Payload Operation \& Application Center (POAC) of TG-2 according to the data transmission schedule every day. The raw data from POLAR are finally transferred from POAC to IHEP through FTP and synchronized from IHEP to the cooperating institutes in Switzerland by rsync\footnote{\url{https://rsync.samba.org}}. All the raw data generated by POLAR directly downloaded from TG-2 are called the 0B level data, all of which are binary data. The 0B level data of POLAR have three types which are defined as SCI\_0B, AUX\_0B and ENG\_0B respectively and they are stored separately in different files. The 0B level data files are right at the start and the initial input of the data pre-processing and scientific data products generation pipeline of POLAR. The three types of 0B level data are firstly summarized in Table~\ref{tab:raw_data}, then described in detail in the subsections. 

\begin{table}[!ht]
\centering
\caption{Summary of raw data from POLAR}\label{tab:raw_data}
\begin{tabular}{|p{1.2cm}|p{4.5cm}|p{2.5cm}|p{4.5cm}|}\hline
Type & Packet Contents & Packet Frequency & Processing Requirements \\\hline
SCI\_0B & Detection data of physical and pedestal events: trigger bits, energy deposition, timestamp, etc. & depending on event rate & Decoding, time alignment between trigger packets and module packets, absolute time reconstruction, etc.  \\\hline
AUX\_0B & Operation mode, threshold and high voltage setting of each module, temperature of each module, etc. & 1 packet every 2 seconds & Decoding, connecting odd packets and even packets, conversion from digital value to physical value, etc. \\\hline
ENG\_0B & Platform parameters data (PPD) of TG-2, part of the housekeeping data, feedback of command injection, digital telemetry data & 1 packet per second for PPD & Extraction of PPD, decoding, coordinate calculation and transformation for the pointing direction of POLAR, etc.\\\hline
\end{tabular}
\end{table}

\subsection{SCI\_0B Data}
All the scientific detection data of POLAR are stored in raw data files of type SCI\_0B. The majority of the scientific detection data is the physical event data collected from the MAPMT. This mainly includes the energy value of each channel with the unit of Analog-to-Digital Converter (ADC) channel, the trigger information of each channel, the time information of each event which is the local time of the instrument, etc. As discussed in \citealt{Produit2017}, POLAR consists of 25 standalone modules with their own FEE at the bottom side and a CT under the OBOX aluminum frame. For each physical event, there will be a trigger packet generated by the CT and one or more module packets generated by the corresponding triggered modules which have channels with energy deposition higher than the hardware threshold. The trigger packet has the information on which modules are being triggered and therefore are having data packets generated, while the module packets contain the information on which channels are being triggered. As each detected physical event has multiple data packets from both CT and FEEs and those packets are stored separately in the raw data file, there is a requirement to align those data packets that belong to the same event  after the packet iterating and decoding. The alignment between the trigger packet from CT and the module packets from FEEs is the basis for the event reconstruction from the raw data of POLAR. The algorithm that performs this alignment for the scientific detection data of POLAR will be presented in Section~\ref{sec:time_alignment}.

\subsection{AUX\_0B Data}
All the engineering data related to POLAR OBOX (also called auxiliary data or housekeeping data), such as the high voltage setting of each MAPMT, the temperature of each module measured by the temperature sensor mounted on the FEE, the OBOX operation mode, etc., are stored in raw data files of type AUX\_0B. Those engineering parameters are very important and necessary for the science data analysis, especially for in-orbit calibration. The raw data of type AUX\_0B also contain the engineering data like the command injection feedbacks, the history of executed commands, the current and voltage measurements of electronics, etc. which are only needed for instrument testing and monitoring but not directly needed by later science data analysis. Unlike type SCI\_0B whose data rate is dependent on the physical event rate, the data rate of the engineering data from POLAR OBOX is fixed, which is one new engineering data packet from OBOX every 2 seconds. Besides the high voltage setting and temperature measurement, there is another kind of very important information stored in the engineering data of OBOX, that is the matching between the local time value of CT and the absolute GPS time value provided by the Pulse Per Second (PPS) signal from the GPS receiver, which is updated every minute by the synchronization (SYNC) command from IBOX. This matching between the local time and the absolute GPS time is used as the reference to reconstruct the absolute time for each physical event that is stored in the raw data of type SCI\_0B. Section~\ref{sec:abs_time_recon} will discuss the method to reconstruct the absolute time for each event. While for the engineering data packets, the absolute time does not need to be calculated separately because the absolute GPS time value has already been tagged into each engineering data packet when it is generated.

\subsection{ENG\_0B Data}
The raw data of both type SCI\_0B and type AUX\_0B are transferred by the Low-Voltage Differential Signaling (LVDS) bus communication while the raw data of type ENG\_0B are transferred by the separated 1553B communication channel from POLAR IBOX. The raw data of type ENG\_0B are the data stream of fixed length (76 bytes) binary packets. Several different types of data are stored in the raw data of type ENG\_0B, which are part of the housekeeping data (like the trigger rate of each module) from POLAR OBOX, the feedback of command injections, the feedback of platform parameters and the digital telemetry data. The four types of data are stored in different packets and can be identified from each other by the special code stored in the packet header. The most important data packet is the one containing the feedback of platform parameters, which is the only necessary data from the raw data of type ENG\_0B that are needed by the science data analysis. Therefore, only the data packets containing the platform parameters are extracted and decoded. One new data packet with the updated platform parameters is generated every second. The platform parameters mainly include the position/velocity in the World Geodetic System 1984 (WGS-84) coordinate system and the attitude information (yaw/roll/pitch angle) of TG-2 platform as well as the corresponding UTC time. These parameters are necessary for calculating the pointing direction of POLAR. The pointing direction of POLAR is necessary for the subsequent calculation of the incident angle of GRBs in POLAR's local coordinate system. The incident angle is needed by both the study of localization of a GRB using the data of POLAR~(\citealt{SUAREZGARCIA2010624}) and the Monte Carlo simulation~(\citealt{Kole2017}) for polarization analysis. The pointing direction of POLAR is first calculated in the WGS-84 coordinate system then transferred to celestial coordinates (J2000) which are widely used in the field of astrophysics. Section~\ref{sec:coo_trans} will discuss the method of the pointing direction calculation and the associated coordinate transformation for POLAR.

\section{Raw data pre-processing pipeline}

\begin{figure}[!ht]
\centering
\includegraphics[width=10cm]{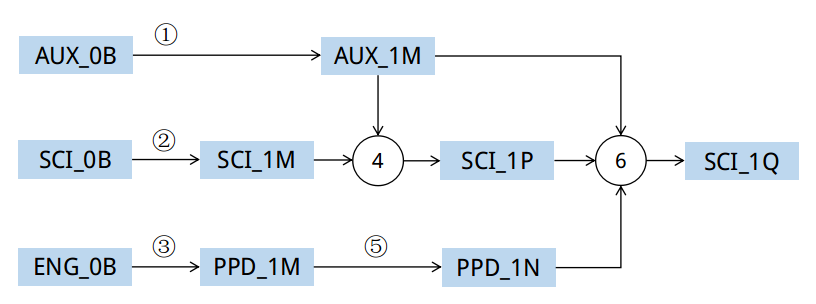}
\caption{Raw data pre-processing pipeline for POLAR}\label{fig:preprocessing_pipeline}
\end{figure}

The pipeline for pre-processing the raw data of POLAR is schematically illustrated by Figure~\ref{fig:preprocessing_pipeline}. The three types of 0B level raw data are firstly decoded by steps \ding{172}, \ding{173} and \ding{174} respectively. The decoding process for the three different types of data involves several common program modules in charge of iterating the frames and packets within each file and extracting values from each packet according to the format specification of the raw data. Those common program modules will be discussed in Section~\ref{sec:iterator_decoder}. Then several other steps are involved for different purposes after the decoding steps. In this section, a general description of the pipeline for each step will be given first, then the main algorithms involved in those steps as well as additional details will be discussed in the subsections. Finally, a subsection discussing the file organization and automatization of the pipeline will be provided.

\begin{dingautolist}{172}
\item Decoding raw data of type AUX\_0B. In this step all the engineering parameters like temperature and high voltage need to be converted from binary data to meaningful physical values after they are extracted. The decoded data of type AUX\_0B in the first stage are called the data product of level AUX\_1M.
\item Decoding raw data of type SCI\_0B. In this step, packets of the same event from FEEs should be correctly aligned to the corresponding packet from CT (Sect.~\ref{sec:time_alignment}). After they are decoded all the packets belonging to the same event should be properly organized by a particular mechanism (Sect.~\ref{sec:event_org}) in the decoded data file. The decoded data of type SCI\_0B in the first stage are called the data product of level SCI\_1M.
\item Decoding raw data of type ENG\_0B. In this step, packets containing the PPD need to first be identified during their iteration then applied to the decoding process. The calculation for the pointing direction of POLAR as well as the coordinate transformation (Sect.~\ref{sec:coo_trans}) also need to be performed in this step after the platform parameters are extracted and decoded. The decoded PPD from the raw data of type ENG\_0B in the first stage are called the data product of level PPD\_1M.
\item As mentioned in Section~\ref{sec:raw_data}, the time of each event in the data of type SCI\_1M decoded from type SCI\_0B is the local CT time in the POLAR instrument. The data of type AUX\_1M decoded from AUX\_0B contain the matching between the local CT time and the absolute GPS time which is updated every minute. Therefore, this step is to use the local CT time in SCI\_1M and the absolute GPS time reference in AUX\_1M to reconstruct the absolute time of each event (Sect.~\ref{sec:abs_time_recon}). The data product after the absolute time of each event is calculated and what is attached is called the data product of level SCI\_1P.
\item The time range for the data files of type SCI\_0B and AUX\_0B is always the same as well as the corresponding AUX\_1M and SCI\_1M (1P) because they are transferred out of POLAR by the same LVDS bus communication channel. While the time range of the data file of type ENG\_0B as well as the corresponding PPD\_1M is different from that of SCI\_0B and AUX\_0B as a result of the different data transmission channel which is the 1553B communication channel. In order to merge the three different types of data in step \ding{177} conveniently, this step is to align the data time range of PPD\_1M with that of SCI\_1P and AUX\_1M by doing the data splitting and merging on the data file of type PPD\_1M. Then the generated data of type PPD\_1N have the same time range as that of SCI\_1P and AUX\_1M.
\item This step merges the three types of data, SCI\_1P, AUX\_1M and PPD\_1N, into a single data file with a simpler data structure (Sect.~\ref{sec:data_merging}), where the organization of event data is highly simplified and each event is attached with the engineering parameters (temperature measurement, high voltage setting, etc.) and the platform parameters (position of TG-2, pointing direction of POLAR, etc.). In the merging process only the data fields from the three types that are necessary for the science data analysis are kept. The merged data file is called the data product of level SCI\_1Q, which is proposed to be the input of the science data analysis like the in-orbit calibration and the polarization analysis.
\end{dingautolist}

\subsection{Iterators and Decoder}\label{sec:iterator_decoder}
The raw data of all the three types are actually the sequence of binary frames and packets. Each type of frame and packet has their specific headers at the beginning. For types SCI\_0B and AUX\_0B, the packets also have the Cyclic Redundancy Check (CRC) code at the end. In the middle are the data corresponding to either scientific detection data or engineering data. The headers are used to identify and separate different frames and packets, while the CRC code is used to check if the frames and packets are transferred correctly. For each type of data a separate iterator module is designed to be in charge of iterating the frames and packets sequence and applying CRC to each frame and packet. The responsibilities of the iterator for each data type are listed below. The decoder is designed as a class with several member functions to extract values from the frames and packets provided by the iterators given the range of bytes or bits.

\begin{figure}[!ht]
\centering
\includegraphics[width=14cm]{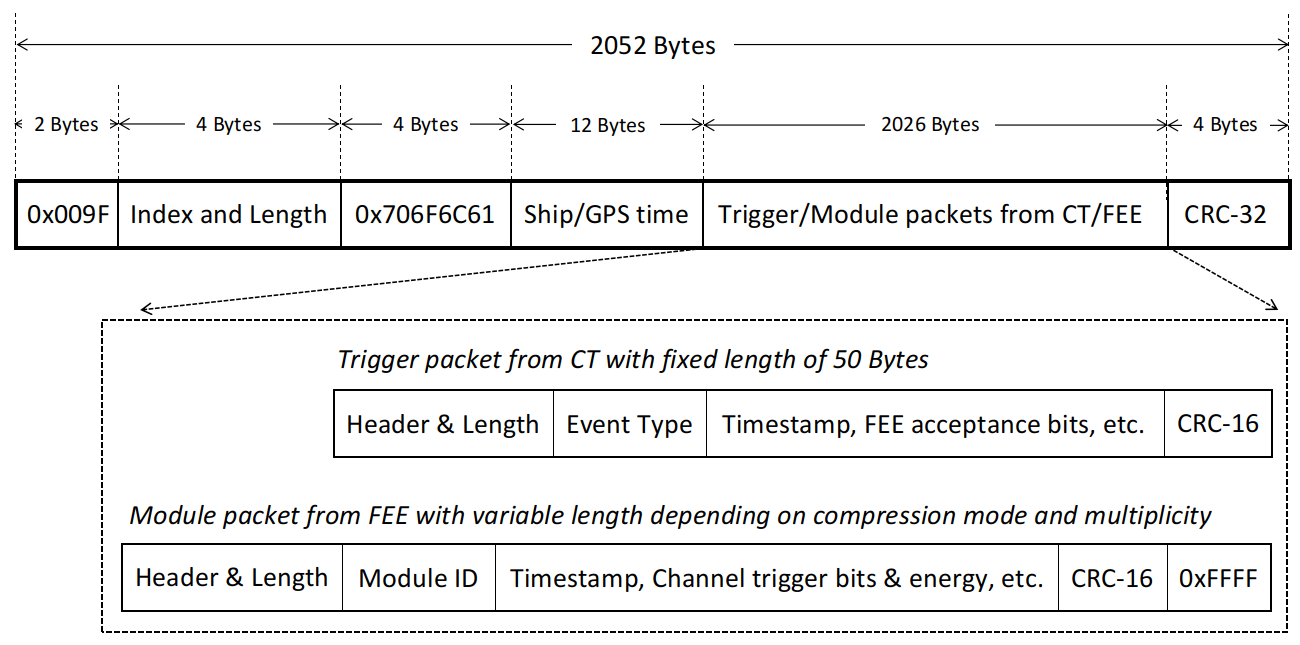}
\caption{Structure of the frame and the packed trigger and module packets of SCI\_0B data.}\label{fig:sciframe}
\end{figure}

\begin{figure}[!ht]
\centering
\includegraphics[width=10cm]{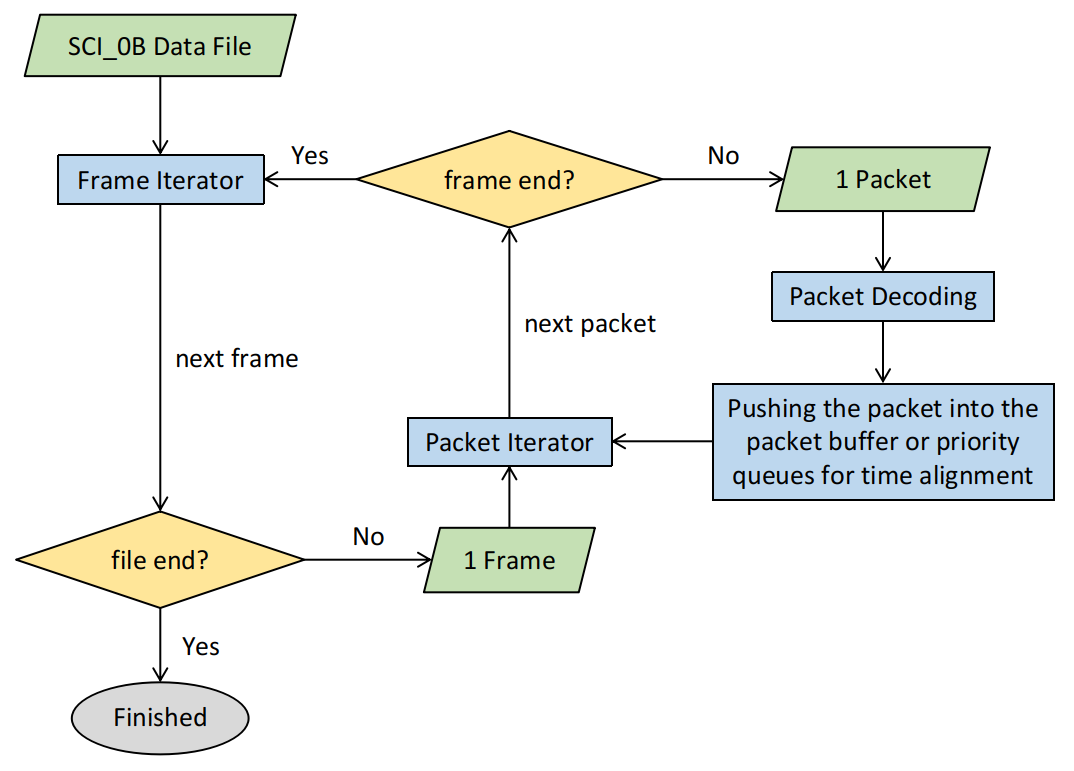}
\caption{Flow chart schematically showing the procedure of frame and packet iterating for SCI\_0B data.}\label{fig:sciframe_iterating}
\end{figure}

\begin{itemize}

\item For the raw data of type SCI\_0B, the data of each physical event consist of one trigger packet from the CT and one or more module packets from the triggered FEEs. Those packets are relatively short and can have variable lengths. Those packets are firstly generated in the OBOX then transferred to the IBOX where they are packed into a series of longer frames with a fixed length of 2052 bytes. The structure of the frame and the trigger and module packets packed in it is shown in Fig.~\ref{fig:sciframe}. The iterator for type SCI\_0B is therefore responsible for iterating the fixed length frames and the short variable length packets within those frames, connecting the packets that are across two adjacent frames and checking the CRC to tag good or bad packets. For SCI\_0B data two different kinds of iterators are designed for frame iterating and packet iterating respectively and the frame and packet iterating procedure is schematically illustrated by the flow chart shown in Fig.~\ref{fig:sciframe_iterating}.

\item For the raw data of type AUX\_0B, each packet of the engineering data from the OBOX is split into two smaller packets in the IBOX which are called the odd packet and the even packet respectively. The length of both is fixed at 260 bytes. These two different packets are transferred out separately from the IBOX every second. The iterator for type AUX\_0B is therefore responsible for iterating the fixed length packets, connecting the odd and even packets which belong to the same OBOX packet and doing the CRC checking.

\item The raw data of type ENG\_0B are simply a sequence of packets with a fixed length of 76 bytes. Therefore the iterator for type ENG\_0B is also very simple, whose responsibility is just to iterate the fixed length packets. During the iteration process, the iterator for type ENG\_0B identifies and only keeps the PPD packets for decoding according to the information in the header as only the PPD data are needed for later science data analysis.

\end{itemize}

\subsection{Time Alignment Algorithm for Physical Events} \label{sec:time_alignment}

The trigger packets and module packets that belong to different events are mixed together in the raw data. The goal of time alignment is to correctly align the trigger packet and the module packets that belong to the same event. The trigger packets from CT and the module packets from the same FEE are always stored in time order, but not guaranteed for the packets from different FEEs and the packets from CT and FEE for different events. The packets of one event have the chance to be transferred out earlier than that of the event before the current event when the physical event rate is high because of the time delay of data transmission between CT and FEE. Therefore, it is hard to only use the information of packet order in the raw data file to align the trigger packet and module packets for physical events.

\begin{figure}[!ht]
\centering
\includegraphics[width=13cm]{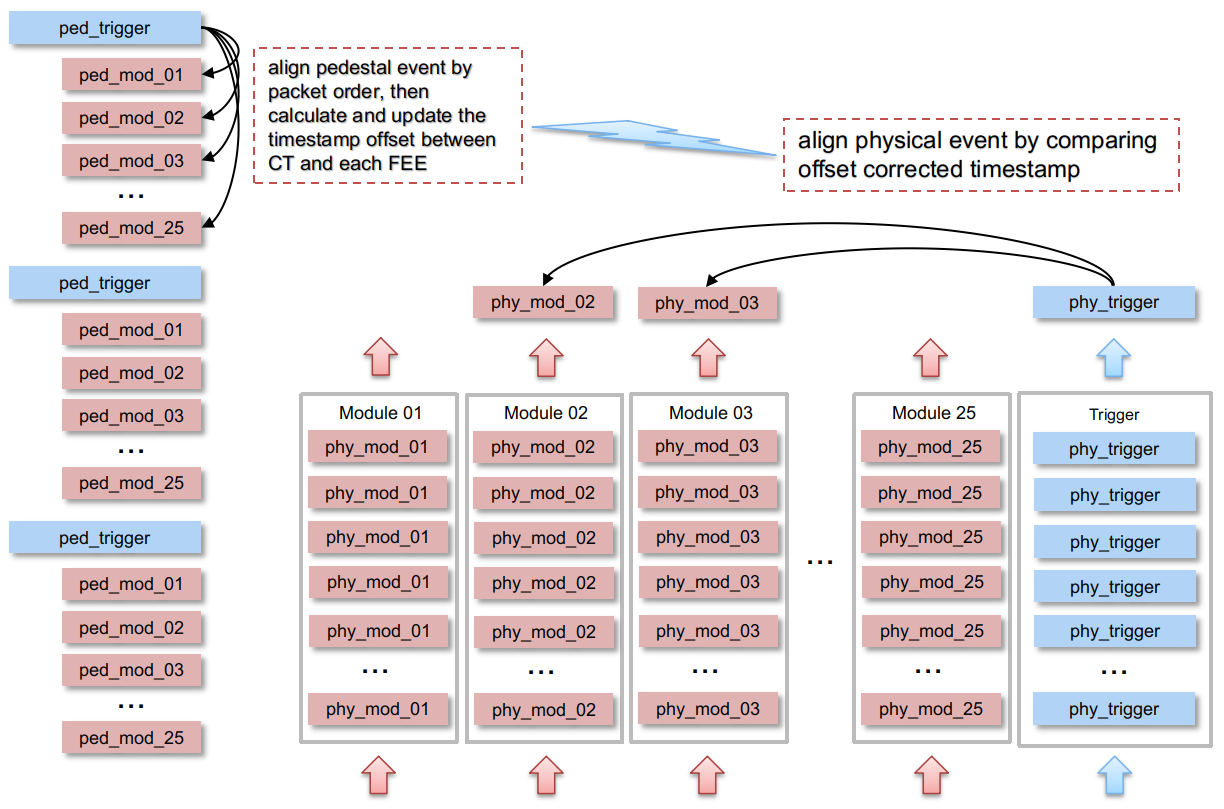}
\caption{Schematic representation of the time alignment algorithm for both pedestal events and physical events. Where blue boxes represent trigger packets from CT, red boxes signify module packets from FEE and gray rectangles mean priority queues. The leftmost column shows the trigger and module packets of pedestal events by the order of packets in the raw data file, and the right columns display the trigger and module packets of physical events after they are pushed into the priority queues.}
\label{fig:time_align}
\end{figure}

Besides the detected physical event data, the raw data of type SCI\_0B also contain the pedestal event data for the purpose of measuring the pedestal level of each channel in the calibration process~(\citealt{LI20188}). Normally each pedestal event has one trigger packet from CT and one module packet from each FEE, which is a total of 26 packets. As the rate of pedestal events is very low (1\,Hz) and the 1 second waiting time between two adjacent pedestal events is far longer than the data transmission time delay, the packets that belong to the same pedestal event are always adjacent to each other in the raw data. It is also found from the raw data that the trigger packet is always transferred out earlier than the module packets for the same event. Therefore, unlike the case of physical events, for pedestal events it is very easy to align the trigger packets and module packets that belong to the same pedestal event only according to the order of the packets in the raw data as shown by the leftmost column in Figure~\ref{fig:time_align}. After the packets of pedestal events are properly aligned, the aligned pedestal events can provide one kind of important information, that is the timestamp offset between CT and FEE. The timestamp offset is caused by the different starting times between CT and FEE. This timestamp offset is needed by the alignment of trigger packets and module packets for physical events based on comparison of timestamps.

For aligning the packets of physical events, two kinds of information can be used. One is the timestamp of the two kinds of packets as mentioned above and the other one is the trigger information about the module that is stored in the trigger packet of each event. The timestamp value after the correction of the offset between CT and FEE can be used to check if two packets belong to the same physical event. The trigger information of the module that is stored in the associated trigger packet is one bit for each module to identify if the module is triggered and therefore has a module packet or not in this event. With the two kinds of information, the alignment algorithm for physical events is implemented as below:

\begin{itemize}
\item Step 1. Establish 25 priority queues\footnote{\url{https://en.wikipedia.org/wiki/Priority_queue}} for the module packets from the 25 FEEs and 1 priority queue for the trigger packet from the CT. In each priority queue, the priority level is set using the timestamp, then all the packets in each priority queue are sorted by time where the packet with the earliest timestamp will always be at the top of the priority queue.
\item Step 2. During the time that the packets are iterating and being decoded, the decoded packets are pushed into the corresponding priority queues according to the type and module ID of the packet.
\item Step 3. As depicted in Fig.~\ref{fig:time_align}, take the trigger packet at the top of the priority queue of trigger packets, then check the trigger information of the module that is stored in the trigger packet and accordingly select the module packets at the top of the priority queues of module packets for only the modules which are triggered in this event. Lastly, compare the timestamps of those module packets after the offset correction with that of the trigger packet to check if those module packets have timestamps that are similar to that of the trigger packet and therefore belong to this event.
\end{itemize}

In this method, the situation of packet loss can also be properly coped with according to the difference in the timestamps between trigger packets and module packets. For example, in Step 3 if the timestamp of one module packet is much later than that of the trigger packet, this can be caused by the loss of the packets by the module, and the module packet with the later (or larger) timestamp should be matched to one of the events after the current event. Therefore, in each trigger packet one extra counter is attached to record the number of module packets that are lost in this event. It should be noted here that in Step 3 the module priority queues should accumulate a certain number of packets before doing the timestamp comparison to avoid an incorrect judgement of packets loss due to the delay of packets transmission between CT and FEEs.

This event alignment method is based on the trigger information of the module that is stored in trigger packets and the timestamp comparison between trigger packets and module packets, therefore this method is also called the time alignment method. According to the processing result using in-orbit data, this method can correctly align more than 99.99\% of all the module packets. This is calculated by computing the ratio of the number of module packets which successfully found their corresponding trigger packet by using the priority queue method and comparing timestamps over the total number of module packets for physical events.

\subsection{Organization of Event Data}\label{sec:event_org}
As mentioned above, the relationship between the trigger packet and module packet for one event is a one-to-many relationship. However the data structures of the two kinds of packets are different and therefore it is hard to store them in the same TTree, which is the main structure to organize data in ROOT. The decoded data in the first stage should contains all the information in the raw data and in order to save data space and to efficiently read the decoded data event by event a special structure is designed to organize the event data in the ROOT file.

\begin{table}[!ht]
\centering
\caption{Example of the special numbers for organizing event data using two TTrees.}
\label{tab:org-event-data}
\scriptsize
\begin{tabular}{| c | c | c | c !{\vrule\,\vrule} c | c | c | c |}\hline
  \multicolumn{4}{| c !{\vrule\,\vrule}}{\textbf{t\_trigger}} & \multicolumn{4}{ c |}{\textbf{t\_modules}} \\\hline
  entry & trigg\_num & pkt\_start & pkt\_count & entry & trigg\_num & ct\_num & event\_num \\\hline
  0 & 0 & 0  & 2 & 0  & 0 & 2  & 0 \\\hline
    &   &    &   & 1  & 0 & 3  & 0 \\\hline
  1 & 1 & 2  & 3 & 2  & 1 & 8  & 0 \\\hline
    &   &    &   & 3  & 1 & 7  & 0 \\\hline
    &   &    &   & 4  & 1 & 2  & 1 \\\hline
  2 & 2 & 5  & 1 & 5  & 2 & 3  & 1 \\\hline
  3 & 3 & 6  & 2 & 6  & 3 & 6  & 0 \\\hline
    &   &    &   & 7  & 3 & 7  & 1 \\\hline
  4 & 4 & 8  & 2 & 8  & 4 & 3  & 2 \\\hline
    &   &    &   & 9  & 4 & 4  & 0 \\\hline
  5 & 5 & 10 & 3 & 10 & 5 & 12 & 0 \\\hline
    &   &    &   & 11 & 5 & 7  & 2 \\\hline
    &   &    &   & 12 & 5 & 8  & 1 \\\hline
  6 & 6 & 13 & 1 & 13 & 6 & 3  & 3 \\\hline
  7 & 7 & 14 & 2 & 14 & 7 & 2  & 2 \\\hline
    &   &    &   & 15 & 7 & 3  & 4 \\\hline
\end{tabular}
\end{table}

As shown in Table~\ref{tab:org-event-data}, two different TTrees are used to store the trigger packets and module packets, the names of which are ``t\_trigger'' and ``t\_modules'' respectively. In each TTree there are some special numbers to maintain the one-to-many relationship between the trigger packet and module packets that belong to the same event. Every valid trigger packet has been given a sequence number which is called ``trigg\_num''. In t\_modules the module packets that belong to the same event are stored adjacently and each module packet records the trigg\_num of the trigger packet for the event. As TTree supports randomly accessing an entry by the entry number and in order to efficiently read the data event by event, each trigger packet in t\_trigger records the position of the corresponding module packets in t\_modules using ``pkt\_start'' and ``pkt\_count'', where pkt\_start records the starting entry number of the sequence of module packets stored in t\_modules and pkt\_count records the number of module packets for the event. The ``ct\_num'' in the module packet is the ID of the module to which the packet belongs, and the ``event\_num'' is a sequence number like the the trigg\_num but the scope of event\_num is within one single module. Table~\ref{tab:org-event-data} lists some examples of these special numbers for organizing event data using two different TTrees, where each non-empty row means an entry in the TTree.

It should be noted here that this structure of event data organization with two different TTrees is only used for the decoded data after the first stage. This structure is to maintain the relationship between trigger packets and module packets and at the same time to reserve all the information in the raw data using a scheme that can save data space as much as possible. Afterwards, a much simpler structure to organize the event data will be introduced for the convenience of science data analysis, which will be discussed in Section~\ref{sec:data_merging}.

\subsection{Absolute Time Reconstruction}\label{sec:abs_time_recon}

\begin{figure}[!ht]
\centering
\includegraphics[width=11cm]{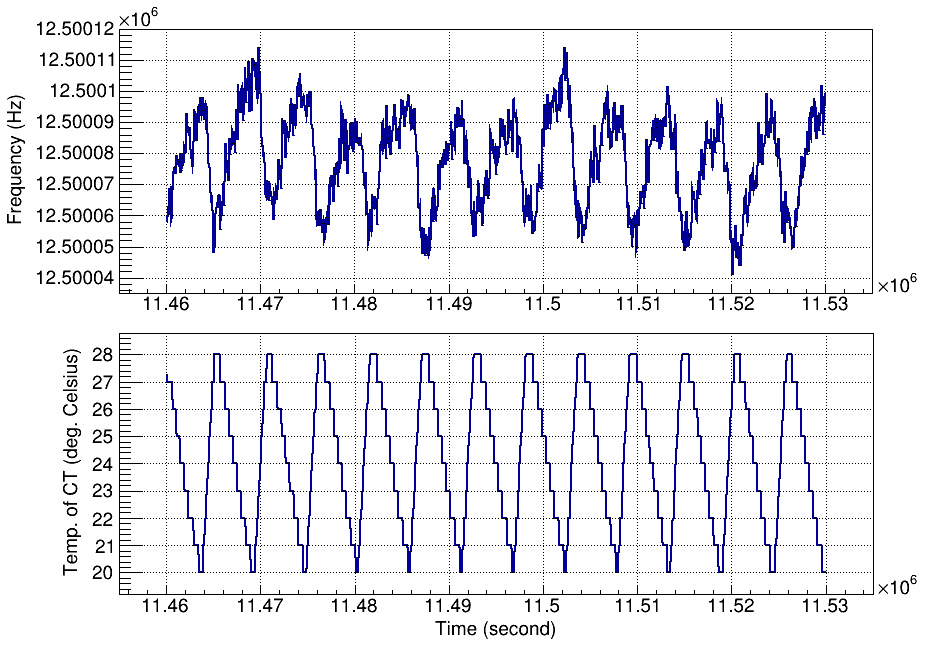}
\caption{The curves representing the frequency of the clock in CT (upper) and the temperature of CT (lower) vs. time.}\label{fig:freq_vs_temp}
\end{figure}

The engineering data of type AUX\_0B contains the matching between the GPS time from PPS and the local timestamp of CT. This matching is updated every minute. In order to reconstruct the absolute time of each event, an iterator to read a pair of GPS time and CT timestamp from the decoded AUX data (AUX\_1M) one by one is designed as the first step. Then when calculating the absolute time of one event, the GPS-timestamp pair, that is nearest in time to the frame in which the trigger packet of the event is located, can be used. The header of each frame in the raw data of SCI\_0B has the GPS time when the frame is generated as shown in Fig.~\ref{fig:sciframe}, therefore it is very easy to calculate the time distance between the frame and the GPS-timestamp pair. Using the GPS-timestamp pair that is nearest in time to the event is mainly concerning the effect of the temperature change on the frequency of the clock in CT. The temperature of the full instrument as well as that of the CT changes periodically due to the position change of the instrument in different orbits. As shown in Figure~\ref{fig:freq_vs_temp}, the frequency of the clock in CT also changes periodically with the change of the temperature of CT. It can also be seen that the frequency of the clock in CT has a clear inverse correlation with the temperature of CT. As the timestamp of CT is given by the clock in CT, in the plot the frequency of the clock in CT is calculated using the two adjacent GPS-timestamp pairs by equation $f = (t_2 - t_1) / (T_2 - T_1)$, where $f$ is the clock frequency, $t_2$ and $t_1$ are the two timestamps and $T_2$ and $T_1$ are the two GPS times in the two adjacent GPS-timestamp pairs. During the process of absolute time reconstruction, the correction for the effect of temperature on the clock frequency is therefore performed for all the events within every minute that are between two adjacent GPS-timestamp pairs. The correction uses the calculated clock frequency every minute instead of the standard frequency of the clock in CT (12.5\,MHz) when computing the absolute time for each event. With the calculated clock frequency $f$, the absolute time of each event is computed by equation $t_{abs} = (t_{loc} - t_1) / f + T_1$, where $t_{loc}$ is the local timestamp of the event and $t_{abs}$ is the reconstructed absolute GPS time of the event in the unit of second.

Using the absolute time of each event reconstructed with the GPS-timestamp pairs and taking the temperature effect into account, the study of evolution of the spin frequency for the Crab pulsar as presented in \citealt{Zheng2017} shows that the spin frequency evolution result for the Crab pulsar achieved using POLAR's data is consistent with that using \textit{Fermi}'s data, which means that the time system of POLAR is accurate and stable. \citealt{Zheng2017} also demonstrates that the sigma of the time residuals for the Crab pulsar observed by POLAR is $84\mathrm{\mu s}$, which includes both the error in the reconstructed absolute time for POLAR and also the non-zero red noise of the spin frequency for the Crab pulsar. This means the error of the reconstructed absolute time for POLAR is less than $84\mathrm{\mu s}$.

\subsection{Pointing Direction Calculation and Coordinate Transformation} \label{sec:coo_trans}
The Z and X axes of POLAR are two vectors that are initially defined in the coordinates of TG-2 after POLAR was mounted. The direction of the Z axis is along while the X axis is perpendicular to the PS bar direction. The directions of the Z and X axes for the POLAR detector need to be calculated in J2000 coordinates. Using the directions of the Z and X axes of POLAR together with the direction of one GRB in J2000 coordinates, the incident angle of the GRB in the detector coordinates can be calculated. The platform parameters that are needed for calculating the directions of the Z and X axes for POLAR in the J2000 coordinate system consist of:
\begin{itemize}
\item the attitude parameters of TG-2, that are the three Euler angles of yaw, roll and pitch of TG-2, which are referred to as $\Psi$, $\phi$ and $\theta$ respectively,
\item the position and instantaneous velocity of TG-2 in the WGS-84  coordinate system, which are respectively referred to as $\bm{P} = (x_w, y_w, z_w)$ and $\bm{V} = (\dot{x}_w, \dot{y}_w, \dot{z}_w)$.
\item the Coordinated Universal Time (UTC) corresponding to the recorded platform parameters, which is referred to as $t$,
\item the directions of Z and X axes in the Guidance, Navigation and Control (GNC) coordinate system of TG-2, which are referred to as $\bm{k}_g = (x_{k}, y_{k}, z_{k})$ and $\bm{i}_g = (x_i, y_i, z_i)$ respectively.
\end{itemize}

According to the coordinate definitions, the direction of the Z (or X) axis is firstly transformed from the GNC coordinate system ($\bm{k}_g$) to the orbit coordinate system ($\bm{k}_o$) by Eq.~\eqref{equ:gnc_to_orbit}.
\begin{equation}\label{equ:gnc_to_orbit}
\bm{k}_o = \bm{R}_{go}(\Psi, \phi, \theta) \cdot \bm{k}_g
\end{equation}
where $\bm{R}_{go}(\Psi, \phi, \theta)$ is the rotation matrix with the rotating order $Z(\Psi) \rightarrow X(\phi) \rightarrow Y(\theta)$\footnote{\url{https://en.wikipedia.org/wiki/Euler\_angles\#Rotation\_matrix}}. Then the direction of the Z (or X) axis is transferred from the orbit coordinates ($\bm{k}_o$) to the WGS-84 coordinates ($\bm{k}_w$) by Eq.~\eqref{equ:orbit_to_wgs84}.
\begin{equation}\label{equ:orbit_to_wgs84}
\bm{k}_w = \bm{R}_{ow}(\bm{P}, \bm{V}) \cdot \bm{k}_o
\end{equation}
where $\bm{R}_{ow}(\bm{P}, \bm{V})$ is the coordinate transformation matrix between the orbit coordinates and the WGS-84 coordinates, which can be calculated using the parameters $\bm{P}$ and $\bm{V}$ according to the definition of the orbit coordinate. Finally, together with the UTC, the direction of the Z (or X) axis in the WGS-84 coordinates ($\bm{k}_w$) needs to be transformed to that in the J2000 coordinates ($\bm{k}_j$). For this transformation, the Python module PyEphem\footnote{\url{http://rhodesmill.org/pyephem}} or the C++ library libnova\footnote{\url{http://libnova.sourceforge.net}} can be used and two basic steps are involved:
\begin{itemize}
\item Step1. calculate the longitude and latitude corresponding to $\bm{k}_w$ and the sidereal time at the current position ($\bm{k}_w$) and UTC (t),
\item Step2. convert the position of $\bm{k}_w$ in the celestial coordinates of the current epoch to that of epoch J2000.
\end{itemize}

With the calculated directions of the Z and X axes of POLAR in the J2000 coordinates and the locations of GRBs provided by other instruments like \textit{Swift} and \textit{Fermi} the computed incident angles of GRBs are as expectated by examining the relative count rates of the 25 modules of POLAR for the detected GRBs. The study of the localization of GRBs using POLAR's data is currently ongoing and the localization results of POLAR are consistent with those provided by other instruments within statistical error. This indicates that the directions of the Z and X axes for POLAR in the J2000 coordinates calculated by the procedure presented in this section are correct.

\subsection{Data Merging and Reorganization}\label{sec:data_merging}

Before the data merging process will be discussed in this section, the event data, the engineering data and the PPD are stored separately in three different files. The structure for organizing the event data is relatively complicated as discussed in Section~\ref{sec:event_org}. This multi-file scheme is not convenient for science data analysis. Therefore, a higher level of data product called SCI\_1Q is generated based on the three kinds of data which are SCI\_1P, AUX\_1M and PPD\_1N. To generate the data product of level SCI\_1Q, the following operations are performed:

\begin{table}[!ht]
\centering
\small
\caption{Some important data fields in the data of level SCI\_1Q}\label{tab:sci_1q}
\begin{tabular}{|l|l|l|}\hline
Type\tablefootnote{\url{https://root.cern.ch/root/html528/ListOfTypes.html}} & Name & Description \\\hline\hline
Long64\_t & event\_id & local sequence number of the event \\\hline
Double\_t & event\_time & MET time of the event\\\hline
Int\_t & type & in-orbit recognized type of the event\\\hline
Bool\_t & is\_ped & indicates if the event is a pedestal event \\\hline
Bool\_t[25] & trig\_accepted & triggering status of each module \\\hline
Bool\_t[25] & time\_aligned & time alignment status of each module \\\hline
Int\_t & pkt\_count & number of module packets for the event \\\hline
Int\_t & lost\_count & number of lost mod. packets for the event \\\hline
Bool\_t[25][64] & trigger\_bit & triggering status of each channel \\\hline
Int\_t & trigger\_n & total number of triggered channels \\\hline
Int\_t[25] & multiplicity & number of triggered channels for each module \\\hline
Float\_t[25][64] & energy\_value & energy deposition value of each channel \\\hline
UShort\_t[25][64] & channel\_status & energy readout status of each channel \\\hline
Float\_t[25] & common\_noise & common noise of each module \\\hline
Int\_t[25] & compress & data compress mode of each module \\\hline
Float\_t[25] & fe\_hv & HV setting of each FEE (from AUX) \\\hline
Float\_t[25] & fe\_temp & temperature of each FEE (from AUX) \\\hline
Double\_t[3] & wgs84\_xyz & position of TG-2 (from PPD) \\\hline
Double\_t[2] & det\_z\_radec & direction of POLAR Z axis (from PPD) \\\hline
Double\_t[2] & det\_x\_radec & direction of POLAR X axis (from PPD) \\\hline
\end{tabular}
\end{table}

\begin{itemize}
\item The trigger packet and module packets of each event are merged into a single TTree, and the pedestal events and physical events are also merged together into the same TTree which are previously stored separately in data level SCI\_1P.
\item The time of each event is converted from GPS time to MET Mission Elapsed Time (MET) starting from the launch time of TG-2. MET is a double-precision floating point value with the unit of second.
\item The engineering data and the platform parameters data that are needed for science analysis are attached to each event according to time. The values of the engineering data and PPD are linearly interpolated in time for each event.
\item In the merging process only the data fields that are directly needed by science analysis are kept, and the data fields that are not necessary for science analysis are discarded.
\item One sub-level number is attached to each data file of level SCI\_1Q to indicate the number of stages that the data file has undergone in the science data analysis pipeline.
\end{itemize}

The data of level SCI\_1Q are designed for the standard science data analysis pipeline as discussed in \citealt{LI20188}. Even though the data size is approximately doubled comparing with SCI\_1P after the merging, with the relatively simple data structure and the single TTree scheme of data level SCI\_1Q, the data analysis pipeline software is highly simplified. Some important data fields in the data of level SCI\_1Q are listed in Table~\ref{tab:sci_1q}.

\subsection{File Organization, Processing Automatization and Email Notification}
When in operation, every day about 40 gigabytes, on average, of 0B level raw data are produced in-orbit by POLAR and transferred to the PSDC in IHEP. The raw data files are firstly organized by a strategy using one day as a unit. The raw data transferred to IHEP on the same day will be collected and permanently stored in a single folder for the day and the folder name is the date of the day. The raw data transferred in the next day will be stored in a new folder. This strategy can safely keep all the old data files intact on the PSDC server when continuously receiving new data files from POAC day by day.

For the high level data products generated by the pipeline shown in Fig.~\ref{fig:preprocessing_pipeline}, the file organization uses a different strategy. All data files with the same type, such as SCI\_1M, AUX\_1M, PPD\_1M, SCI\_1P, PPD\_1N and SCI\_1Q, are stored in the same folder with the type name, no matter which day the files are generated. The name of each data file has the exact time range of the internal data with a precision of 1 second. This strategy is chosen for the convenience of processing in the pipeline's automatization as well as for users easily finding the data files corresponding to a specific type within a given time range.

Manually finding and processing all the new raw data files every day are very time consuming. Therefore a series of scripts written by Python is prepared to automatize the whole data processing pipeline from 0B level to 1Q level as presented by Fig.~\ref{fig:preprocessing_pipeline}. Those scripts automatically start to run at some specific time points every day coordinating with the schedule of data transferring from POAC to IHEP. Once those scripts start to run they will firstly identify the new raw data files in the raw data folder from the current day by comparing file names. Subsequently, they will invoke the data pre-processing programs written by C++ to generate the high level data products for the new raw data and put them into the corresponding folders. For each file of the high level data products, the command to generate it and the screen output as well as the log output are all reserved in specific folders for later checking or regenerating the file when needed. After all the new raw data files are finished, to generate the corresponding high level data products by the automatization scripts, one notification email will be generated based on what has been processed and sent by the built-in SMTP Python client to researchers in the POLAR group who can access the data products on the PSDC server in IHEP. The notification email mainly includes the starting and finishing time for processing the newly found raw data, a list of all the new raw data files that have been processed, a list of all the new corresponding high level data product files that have been generated and information on any errors that occurred during the processing. According to the notification email, the researchers can login the PSDC server and analyze the interesting data files on the server and the staff in charge of maintaining the data products deal with the exceptions and errors found during the processing. With the automatic scripts, the full pipeline from the 0B level raw data to the 1Q level data product as presented by Fig.~\ref{fig:preprocessing_pipeline} can be finished within 3 hours for 10 gigabytes of raw data files after they arrive at IHEP. This is efficient enough for the following science data analysis.

This raw observation data pre-processing and high level scientific data products generation pipeline and the automatic scripts introduced in this section have already continuously worked for several months without many manual interventions on the PSDC server in IHEP for POLAR. The core software units used in this data processing pipeline, such as the decoding and time alignment program for SCI\_0B data, the decoding program for AUX\_0B data, the absolute time reconstruction program, etc., were fully tested and extensively used by processing the ground calibration data and firmware testing data before the launch of POLAR. The full pipeline was validated well by the success of automatically processing all the in-orbit data from POLAR and providing service for the science data analysis of researchers after POLAR was launched.

\section{Higher level Scientific data products}

With the data of level SCI\_1Q, most of the science analysis for POLAR can be performed. However, concerning the future data publication as well as the convenience of the internal use of the data from POLAR, some higher level scientific data products can be designed and generated. Generating the higher level scientific data products mainly consists of two aspects: data archiving and the scientific data products of specific celestial sources. For data archiving, it needs to recombine the data files of level SCI\_1Q in order to remove the data duplication among different files and to organize each data file with a fixed short time length, for example, making each data file have one hour of data. For the scientific data products associated with specific celestial sources, it needs to firstly extract the data segment of SCI\_1Q at the time when one event, like a GRB or solar flare, occurred and was detected by POLAR, then add some property information about the event, such as the T0, T90, location in the J2000 coordinates, incident angle and the energy spectrum parameters of the source. The energy spectrum parameters can be either given by other instruments or provided by the analysis using the data from POLAR itself. As the FITS\footnote{\url{https://fits.gsfc.nasa.gov}} file format is widely used in the field of astronomy, either the FITS files or the tools to convert the ROOT files to FITS files for each type of higher level scientific data products can also be provided when they are published. The process of generating the higher level scientific data products will not involve some complicated procedures besides the simple procedures like data splitting and merging for the data files of SCI\_1Q and the addition of some extra meta-information related to the source, therefore, this section will not give more details about this process.

\section{Summary and Conclusions}
POLAR is a space-borne GRB polarimeter onboard the Chinese space laboratory TG-2 which was successfully launched on 2016 September 15. In order to pre-process the in-orbit data from POLAR that are transferred from TG-2 to the ground as quickly as possible and generate the high level scientific data products that are suitable for later science analysis, a full software pipeline was designed and developed before the launch date of POLAR and this paper presented a detailed introduction to the pipeline. The three types of raw data from POLAR and the pre-processing requirements are firstly introduced. Then each step of the pipeline is discussed as well as the relevant algorithms and methods that are used in those steps. The most important data are the event data detected by the POLAR detector, and the event data are converted by four steps: SCI\_0B $\rightarrow$ SCI\_1M $\rightarrow$ SCI\_1P $\rightarrow$ SCI\_1Q. The scientific data product of level SCI\_1Q simplified the data structure of the event data and merged the necessary engineering data and PPD that are respectively decoded from the raw data of types AUX\_0B and ENG\_0B. The SCI\_1Q data product is chosen as the input and as a standard data format in the science data analysis pipeline in the PSDC at IHEP. Some higher level scientific data products for the data archiving and specific events like GRBs and solar flare events can be generated based on the data product of level SCI\_1Q. The in-orbit observation data pre-processing and scientific data products generation pipeline and the related software can work automatically without manual intervention for most of the time after the raw data from POLAR arrive at IHEP and have been continuously and successfully used for months by the PSDC in IHEP after POLAR was launched and switched on. In practice, the full pipeline from the 0B level raw data to the 1Q level data product for the science data of POLAR can be finished within 3 hours for 10 gigabytes of raw data files after they arrive at IHEP, which is efficient enough for the following science data analysis. Also, after the pipeline for all the newly-arrived raw data has finished, an email describing the generation of new scientific data products will be automatically sent to the researchers who can access the scientific data products. With the scientific data products generated by this pipeline the in-orbit calibration of POLAR has been finished and published so far and the first scientific results of POLAR have been produced and published recently.

\newpage
\begin{acknowledgements}
We gratefully acknowledge financial support from the Joint Research Fund in Astronomy under a cooperative agreement between the National Natural Science Foundation of China and the Chinese Academy of Sciences (Grant No. U1631242), the National Natural Science Foundation of China (Grant No. 11503028, 11403028), the Strategic Priority Research Program of the Chinese Academy of Sciences (Grant No. XDB23040400) and the National Basic Research Program (973 Program) of China (Grant No. 2014CB845800).
\end{acknowledgements}

\bibliographystyle{raa}
\bibliography{references}

\end{document}